\documentclass[12pt,authoryear,3p,times]{elsarticle}

\usepackage{amssymb}
\usepackage{amsmath}
 \usepackage{relsize}
\usepackage{booktabs}
\usepackage{hyperref}
\usepackage[utf8]{inputenc}
\usepackage[T1]{fontenc} 
\usepackage{multirow}
\usepackage[table,xcdraw]{xcolor}
\usepackage[dvipsnames]{xcolor} 
\usepackage{placeins}
\usepackage{orcidlink}

\journal{International Journal of Forecasting}

\begin{document}

\begin{frontmatter}

\renewcommand{\thefootnote}{$\bigstar$}

\title{Stealing accuracy: Predicting day-ahead electricity prices with temporal hierarchy forecasting (THieF)\footnote{Forthcoming in the \textit{International Journal of Forecasting}, 2026}}

\author[ONAS]{Arkadiusz Lipiecki \orcidlink{0000-0003-1118-0388}}
\author[ONAS]{Kaja Bili\'nska \orcidlink{0009-0008-2658-2690}} 
\author[Skovde]{Nikolaos Kourentzes \orcidlink{0000-0003-0211-5218}}
\author[ORBI]{Rafał Weron \orcidlink{0000-0003-1619-5239} \corref{cor1}}
\address[ONAS]{Department of Computational Social Science, Wrocław University of Science and Technology, Poland}
\address[Skovde]{Skövde AI Lab, University of Skövde, Sweden}
\address[ORBI]{Department of Operations Research and Business Intelligence, Wrocław University of Science and Technology, Poland}
\cortext[cor1]{Corresponding author; \textit{email:} rafal.weron@pwr.edu.pl}

\begin{abstract}
We introduce the concept of \textit{temporal hierarchy forecasting} (THieF) in predicting day-ahead electricity prices and show that reconciling  forecasts for hourly products and 2- to 24-hour blocks can significantly (up to 13\%) improve accuracy at all levels. These results remain consistent throughout a challenging 4-year test period (2021-2024) in the German and Spanish power markets and across model architectures, including linear regression, shallow feedforward neural networks, gradient-boosted decision trees, and a state-of-the-art, pretrained transformer. 
Given that (\textit{i}) trading of block products is becoming more common and (\textit{ii}) the computational cost of reconciliation is comparable to that of predicting hourly prices alone, we recommend using it in daily forecasting practice. 
\end{abstract}


\end{frontmatter}

\section{Introduction}

Operational decisions often require tailored short-term forecasts that focus on different levels of detail and granularity \citep{pet:etal:22}. For instance, models for hourly products in wholesale electricity markets can use different information sets than those for average daily prices \citep{wer:14,mac:wer:16}. 
These forecasts may not align, which can lead to suboptimal decisions. To cope with this, the forecasts from each temporal level of the hierarchy should be reconciled to be coherent. 

The last decade has seen an unprecedented growth in interest in forecast reconciliation \cite[see][for a recent review]{ath:hyn:kou:pan:24:rev} and the introduction of \textit{temporal hierarchy forecasting}  \cite[THieF;][]{ath:hyn:kou:pet:17}.
The latter can be applied to any time series by means of non-overlapping temporal aggregation -- the predictions computed at all levels of the hierarchy are combined to yield temporally reconciled, accurate and robust forecasts. 
The concept is new in energy forecasting and there are only a handful of publications on predicting electric load \citep{nys:lin:pin:mad:20,mol:nys:mad:24} or wind \citep{jeo:pan:pet:19,eng:abo:24,sha:bha:jai:24} and solar \citep{yan:qua:dis:r-g:17,dif:gir:23} generation. 
More importantly, temporal hierarchies have not yet been applied to \textit{electricity price forecasting} (EPF). The only related study predicts aggregated demand and supply curves by exploiting their intrinsic hierarchical structure \citep{ghe:zie:24}, but is not concerned with temporal hierarchy forecasting nor EPF.

To fill this gap and provide market participants with a universal tool to improve the predictions of hourly, block and baseload prices (i.e., daily average prices), we conduct an extensive study involving four classes of models and two 4-year test periods (2021-2024) from two major European power markets: Germany and Spain. We use a variety of base forecasting methods with increasing complexity and diverse modeling assumptions, ranging from a parsimonious expert model built on classical linear regression \citep{zie:wer:18,ser:wer:25} to Amazon's state-of-the-art tabular foundation model based on a 12-layer 72 million-parameter transformer architecture \citep{zha:rob:25}.

To our best knowledge, we are the first to use pretrained foundation models not only for EPF, but also for temporal hierarchies. As such, we complement the literature that has focused on evaluating the benefits of THieF with conventional, relatively small, base forecasting methods. 
We find that temporal reconciliation leads to significant improvements in terms of the mean absolute error (MAE) and the root mean squared error (RMSE) across all classes of models and hierarchy levels considered: up to 5.5\% for hourly and 13.4\% for baseload prices.

The remainder of the paper is structured as follows. In Section \ref{sec:THieF} we briefly review the THieF approach and discuss its application to electricity price forecasting. Next, in Section \ref{sec:Datasets} we introduce the datasets and in Section \ref{sec:base:forecasts} we explain how the base forecasts for the 24 hourly prices are determined using four classes of models: linear regression ($\rightarrow$ Section \ref{ssec:ARX}), committee machines of shallow feedforward neural networks ($\rightarrow$ Section \ref{ssec:NARX}), gradient-boosted decision trees ($\rightarrow$ Section \ref{ssec:XGB}), and a state-of-the-art, pretrained transformer ($\rightarrow$ Section \ref{ssec:Mitra}). In Section \ref{sec:Results} we discuss the results obtained. Finally, in Section \ref{sec:Conclusions}, we conclude and speculate on how our approach can be extended in future studies.

\section{Temporal hierarchies and reconciliation}
\label{sec:THieF}

THieF can be understood in three steps: (\textit{i})~construct non-overlapping temporally aggregate time series (levels), (\textit{ii})~generate base forecasts at these levels independently, (\textit{iii})~reconcile the base forecasts to combine the diverse information \citep{ath:hyn:kou:pet:17}. Temporal aggregation is a moving average that filters and strengthens different aspects of the original signal, therefore, the combination of the base forecasts helps recover information that would otherwise be difficult to estimate \citep{kou:pet:tra:14}. 

\begin{figure*}[tbp]
\includegraphics[width=\linewidth]{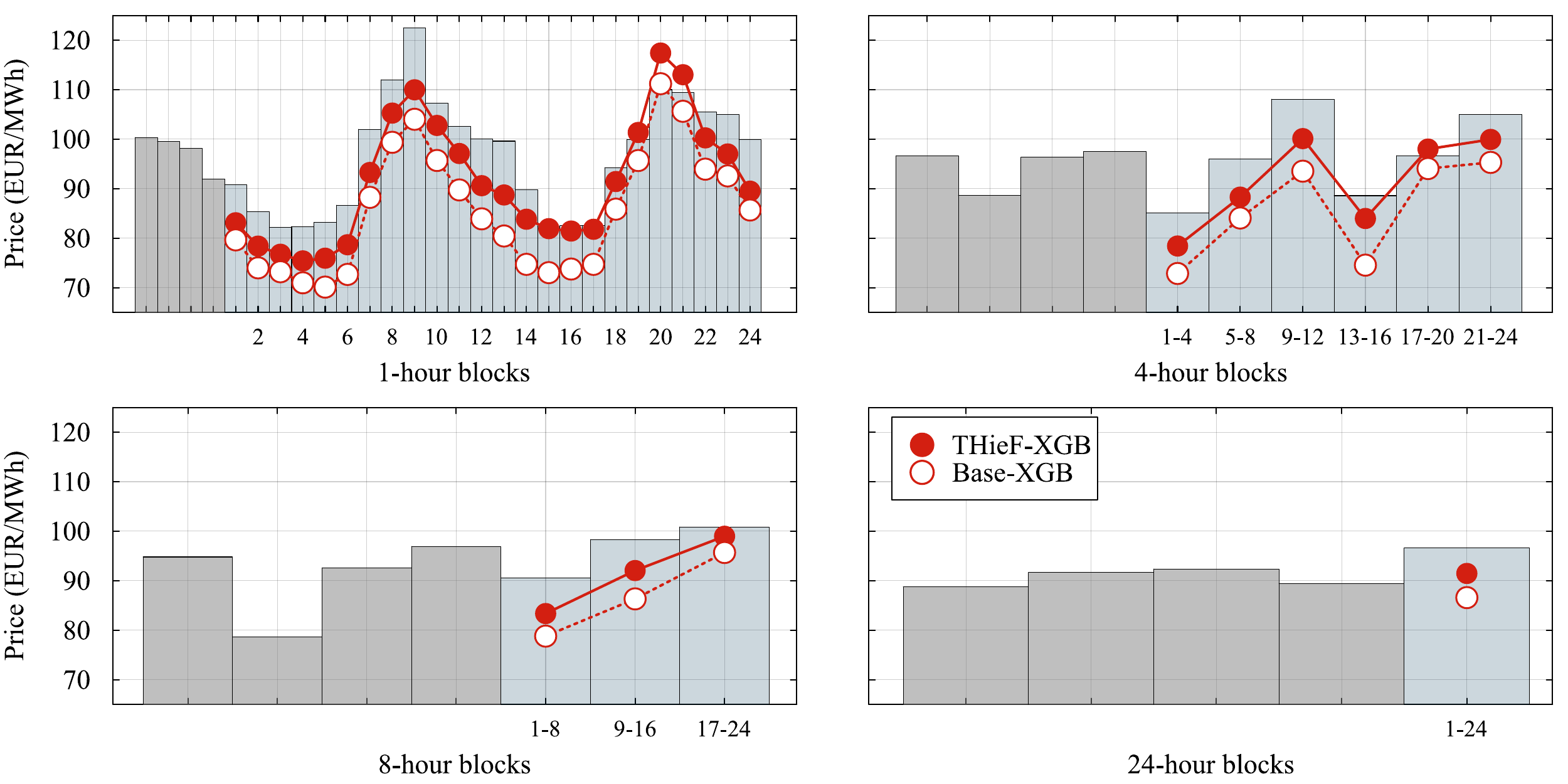}
\caption{Stylized example of the impact of THieF on forecasts at four hierarchy levels (1H, 4H, 8H, 24H). Filled red circles denote reconciled ($\rightarrow$ THieF-XGB) and hollow circles unreconciled forecasts ($\rightarrow$ Base-XGB) obtained using eXtreme Gradient Boosting (see Section \ref{ssec:XGB}). Bars represent actual block prices, with light blue indicating the target day, i.e., Sunday, 7.02.2021, in the German EPEX-DE market.}
\label{fig:THieF}
\end{figure*}

The benefits of THieF primarily stem from three elements. First, it independently models different views of the time series, which is particularly advantageous in the presence of increased model misspecification \citep{ath:hyn:kou:pet:17}. Second, it relies on an implicit forecast combination, leveraging the relationships among temporally aggregated series \citep{kou:ath:19}. Third, it uses restricted estimation of the implied combination weights, which reduces estimation uncertainty \citep{pri:sve:kou:21}.

In Figure~\ref{fig:THieF} we plot the out-of-sample base and reconciled forecasts at four aggregation levels for a sample day, market, and model. THieF blends the information of the base forecasts, and the relatively low bias of the average daily price forecast helps to reduce the bias at all other levels. The combination achieved by THieF is affine, but not convex (weights sum to unity but can be negative). This can help to further improve forecasts.

\begin{figure}[tbp]
\centering
\includegraphics[width=0.5\linewidth]{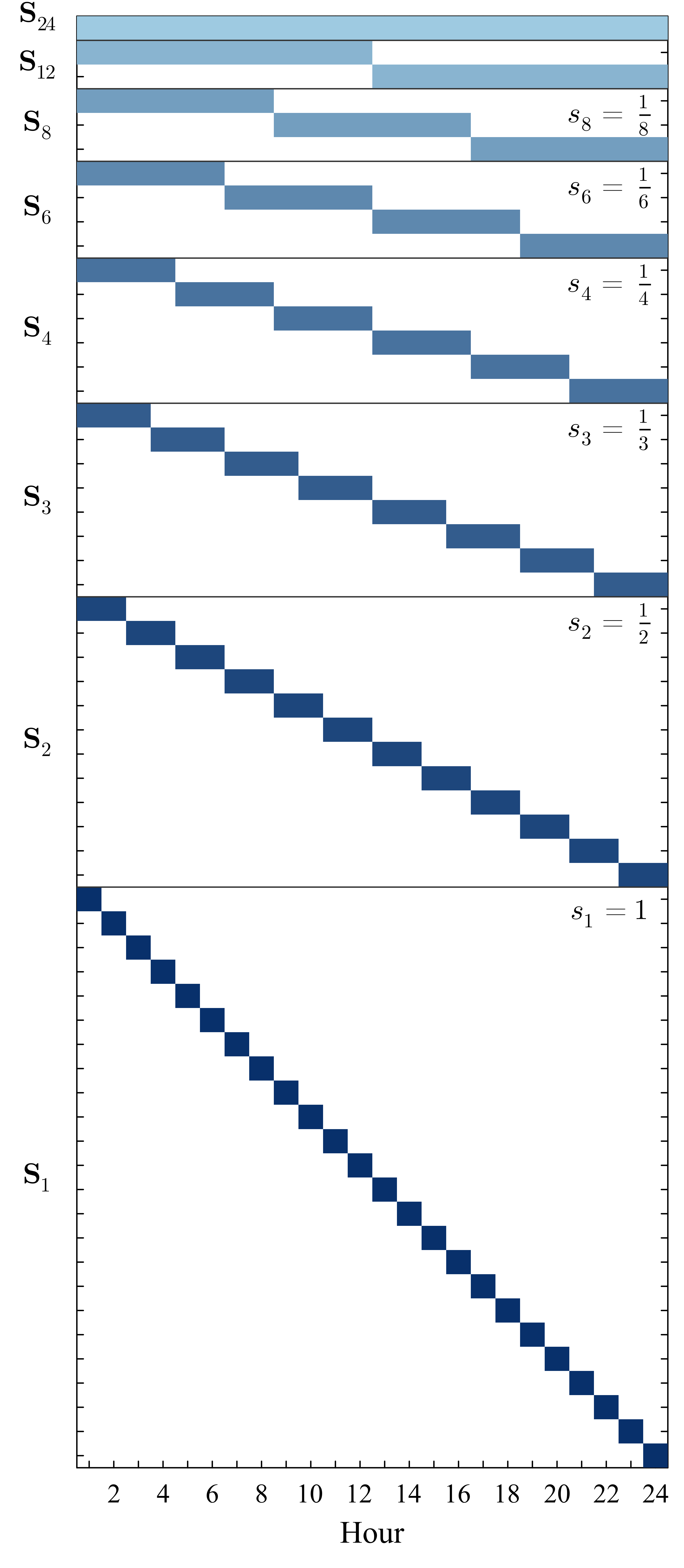}
\caption{The summing matrix $\mathbf{S}$ defined in Eq.~\eqref{eq:summing_matrix_1}. Blue blocks denote non-zero elements, and the shading indicates weights ranging from $s_{24}=\frac{1}{24}$ for $\mathbf{S}_{24}$ to $s_1=1$ for $\mathbf{S}_1$.}
\label{fig:summing_matrix}
\end{figure}

\subsection{The summing matrix}

Let $m$ be the frequency of a time series, for instance, $m=24$ for hourly data. Let $k = \{k_i\}_{i=1}^p$ be the divisors of $m$, ordered such that $m = k_p > k_{p-1} > \ldots > k_1 = 1$, and $p$ is the number of factors. For example, for hourly data we have $k_i \in \{24, 12, 8, 6, 4, 3, 2, 1\}$ and $p=8$. We define the \textit{summing matrix} of size $(\sum_{i=1}^p{k_i}) \times m$ as: 
\begin{equation}\label{eq:Smat}
\mathbf{S} = \begin{bmatrix} \mathbf{S}_{k_p}^T \dots \mathbf{S}_{k_1}^T \end{bmatrix}^T,
\end{equation}
where $\mathbf{S}_{k}^T$ is the transpose of $\mathbf{S}_{k}$. Each $\mathbf{S}_{k}$ is given by:
\begin{equation}\label{eq:S_k}
\mathbf{S}_{k} = s_{k} \mathbf{I}_{m/k} \otimes \mathbf{J}_{1,k},
\end{equation}
i.e., the weight $s_k = \frac{1}{k}$ multiplied by the Kronecker product between the identity matrix $\mathbf{I}_{m/k}$ of size $\frac{m}{k} \times \frac{m}{k}$ and a row vector $\mathbf{J}_{1,k}$ of size $1 \times k$ with all its elements equal to 1. 
For instance, for hourly data we get:
\begin{equation}\label{eq:summing_matrix_1}
    \mathbf{S} = \begin{bmatrix}
        \mathbf{S}_{24}^T \; \mathbf{S}_{12}^T \;  \mathbf{S}_{8}^T  \; \mathbf{S}_{6}^T \; \mathbf{S}_{4}^T \; \mathbf{S}_{3}^T \; \mathbf{S}_{2}^T \;\mathbf{S}_{1}^T
        \end{bmatrix}^T.
\end{equation}
This is illustrated in Figure~\ref{fig:summing_matrix}. The top row $\mathbf{S}_{24}$ represents the baseload block, i.e., the average of the 24 hourly prices. The two rows of $\mathbf{S}_{12}$ correspond to the 12-hour blocks, respectively for hours 1-12 and hours 13-24. Analogously, the three rows of $\mathbf{S}_{8}$ map the hourly prices to the average price for the first, middle and last 8-hourly blocks. The bottom-level $\mathbf{S}_{1}$ is a $24\times24$ identity matrix and corresponds to forecasts for the individual hours.

The temporal aggregation levels prescribed by $k$ are non-overlapping for two reasons. First, overlapping temporal aggregation introduces serial autocorrelation that would need to be captured by the base forecasting methods. This not only increases model complexity but can also obscure identification of the underlying signal. Second, under the reasonable assumption that the base models may be misspecified, such serial autocorrelation can further complicate reconciliation in THieF.

\subsection{Reconciliation}
 
Let $\mathbf{p}_d$ be a column vector that contains the hourly observations for day $d$, then $\mathbf{P}_d=\mathbf{S}\mathbf{p}_d$ implements step (\textit{i}). Further, let $\mathbf{\hat{P}_{d}}$ be a vector of unreconciled base forecasts $\rightarrow$ step (\textit{ii}). Following \cite{ath:hyn:kou:pet:17}, THieF reconciliation is performed each day by computing:
\begin{equation}\label{eq:reconciliation}
    \mathbf{\tilde{P}_{d}} = \mathbf{S}(\mathbf{S}^T \mathbf{W}^{-1}\mathbf{S})^{-1} \mathbf{S}^T \mathbf{W}^{-1} \mathbf{\hat{P}_{d}},
\end{equation}
where $\mathbf{W}$ is the covariance matrix of the base forecast errors, estimated on the errors from the training sample $\rightarrow$ step (\textit{iii}). Since the covariance matrix for our hierarchical forecasts is $60\times60$, the total number of elements to estimate (1800) is larger than the sample size (1085; see Section \ref{sec:Datasets}). Therefore, we resort to regularized covariance matrix estimators.

Note that our formulation differs from \cite{ath:hyn:kou:pet:17} by using the mean for the aggregation, as in \cite{kou:pet:tra:14}. This is due to the effect of weights $s_k$ in the calculation of each $\mathbf{S}_{k}$ in Eq.~\eqref{eq:S_k}. Effectively, this normalizes $\mathbf{W}$.

To see this, let $\mathbf{S}_\Sigma$ be the summing matrix as in \cite{ath:hyn:kou:pet:17}, which is constructed using Eq.~\eqref{eq:Smat}, but with all $s_k \equiv 1$. Further, we set $\mathbf{\Lambda} = diag(1/\mathbf{S}_{\Sigma}\mathbf{1})$ and use it to rescale sum-aggregates into mean-aggregates. Since $\mathbf{S} = \mathbf{\Lambda} \mathbf{S}_{\Sigma}$ and $\mathbf{\Lambda}^T = \mathbf{\Lambda}$, we can rewrite Eq.~\eqref{eq:reconciliation} as:

\begin{equation} 
\mathbf{\tilde{P}_{d}} = \mathbf{\Lambda}\mathbf{S}_{\Sigma} \left( \mathbf{S}_{\Sigma}^T \mathbf{\Lambda} \mathbf{W}^{-1} \mathbf{\Lambda} \mathbf{S}_{\Sigma} \right)^{-1} \mathbf{S}_{\Sigma}^T \mathbf{\Lambda} \mathbf{W}^{-1} \mathbf{\hat{P}_{d}}.
\end{equation}
Given that $\mathbf{W}_{\Sigma}^{-1} = (\mathbf{\Lambda}^{-1} \mathbf{W} \mathbf{\Lambda}^{-1})^{-1} = \mathbf{\Lambda} \mathbf{W}^{-1} \mathbf{\Lambda}$ and multiplying both sides by $\mathbf{\Lambda}^{-1}$ we obtain:
\begin{equation}
\mathbf{\Lambda}^{-1} \mathbf{\tilde{P}_{d}} = \left[ \mathbf{S}_{\Sigma} ( \mathbf{S}_{\Sigma}^T \mathbf{W}_{\Sigma}^{-1} \mathbf{S}_{\Sigma} )^{-1} \mathbf{S}_{\Sigma}^T \mathbf{W}_{\Sigma}^{-1} \right] \mathbf{\Lambda}^{-1} \mathbf{\hat{P}_d}.
\end{equation}
The terms within the square brackets are identical to the reconciliation using $\mathbf{S}_{\Sigma}$ in \cite{ath:hyn:kou:pet:17}. Therefore, the use of $\mathbf{S}$ instead of $\mathbf{S}_{\Sigma}$ normalizes $\mathbf{W}$, but otherwise results in the equivalent sum-aggregate forecasts.

\subsection{Shrinkage}

We considered three shrinkage approaches for estimating the covariance matrix $\mathbf{W}$, the constant correlation model \citep{led:wol:04}, the diagonal shrinkage target \citep{sch:str:2005}, which has been shown to perform well in forecast reconciliation \citep{wic:etal:2019}, and the so-called `variance scaling' \citep{ath:hyn:kou:pet:17} that sets all off-diagonal elements to zero and corresponds to the Weighted Least Squares (WLS) estimator. We found that the two first approaches performed similarly well. The variance scaling also provided improvements, but more mediocre than the alternatives.
In Section \ref{sec:Results}, we report the results for the less conservative, constant correlation shrinkage of \cite{led:wol:04}. The results for the remaining approaches are provided in \ref{app:extraW}.

\begin{figure*}
    \centering
    \begin{minipage}{\linewidth}
        \centering
        Germany (EPEX-DE)
        \includegraphics[width=\linewidth]{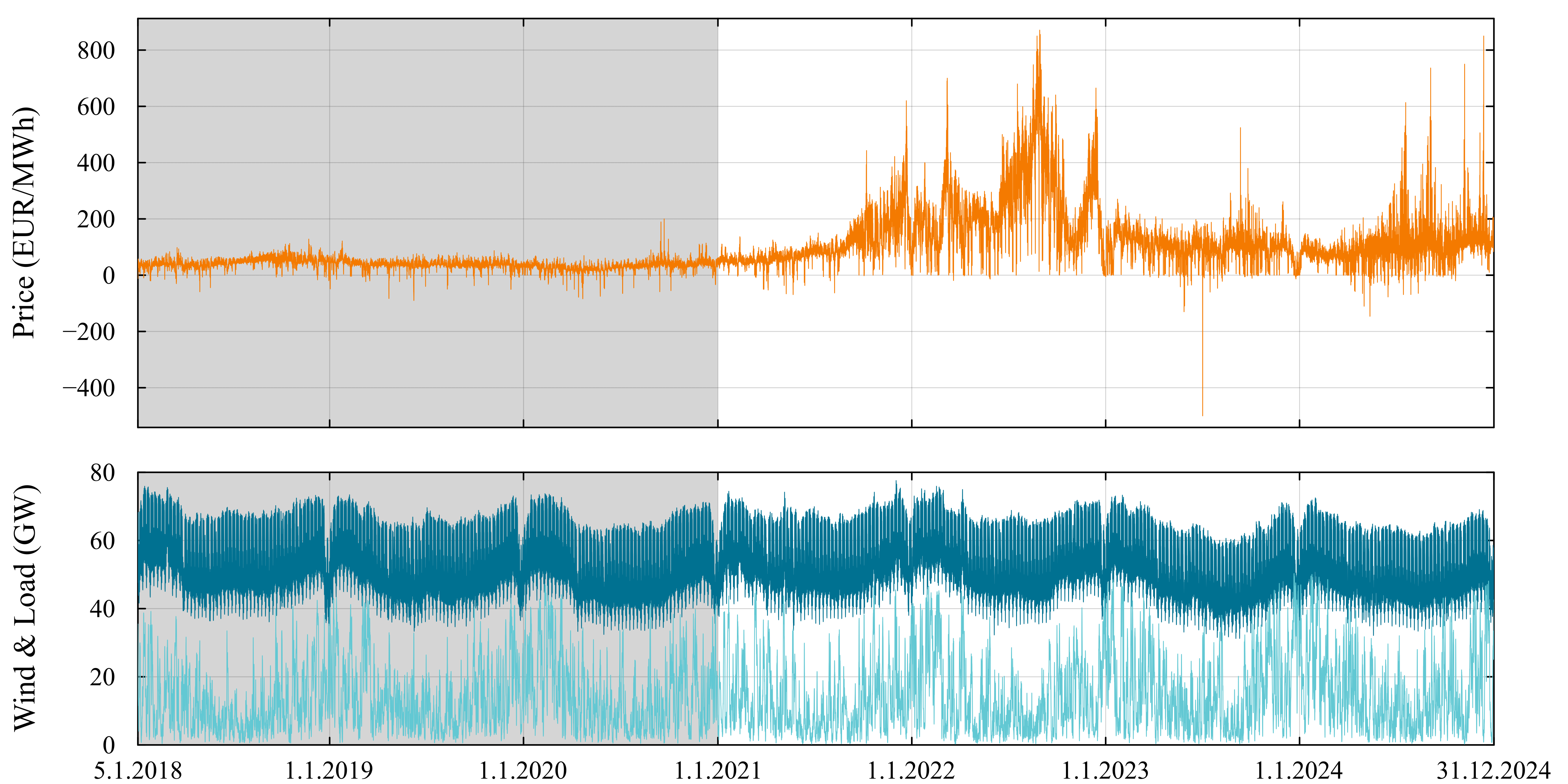}
    \end{minipage}
    \begin{minipage}{\linewidth}
        \centering
        \vspace{0.5em} Spain (OMIE)
        \includegraphics[width=\linewidth]{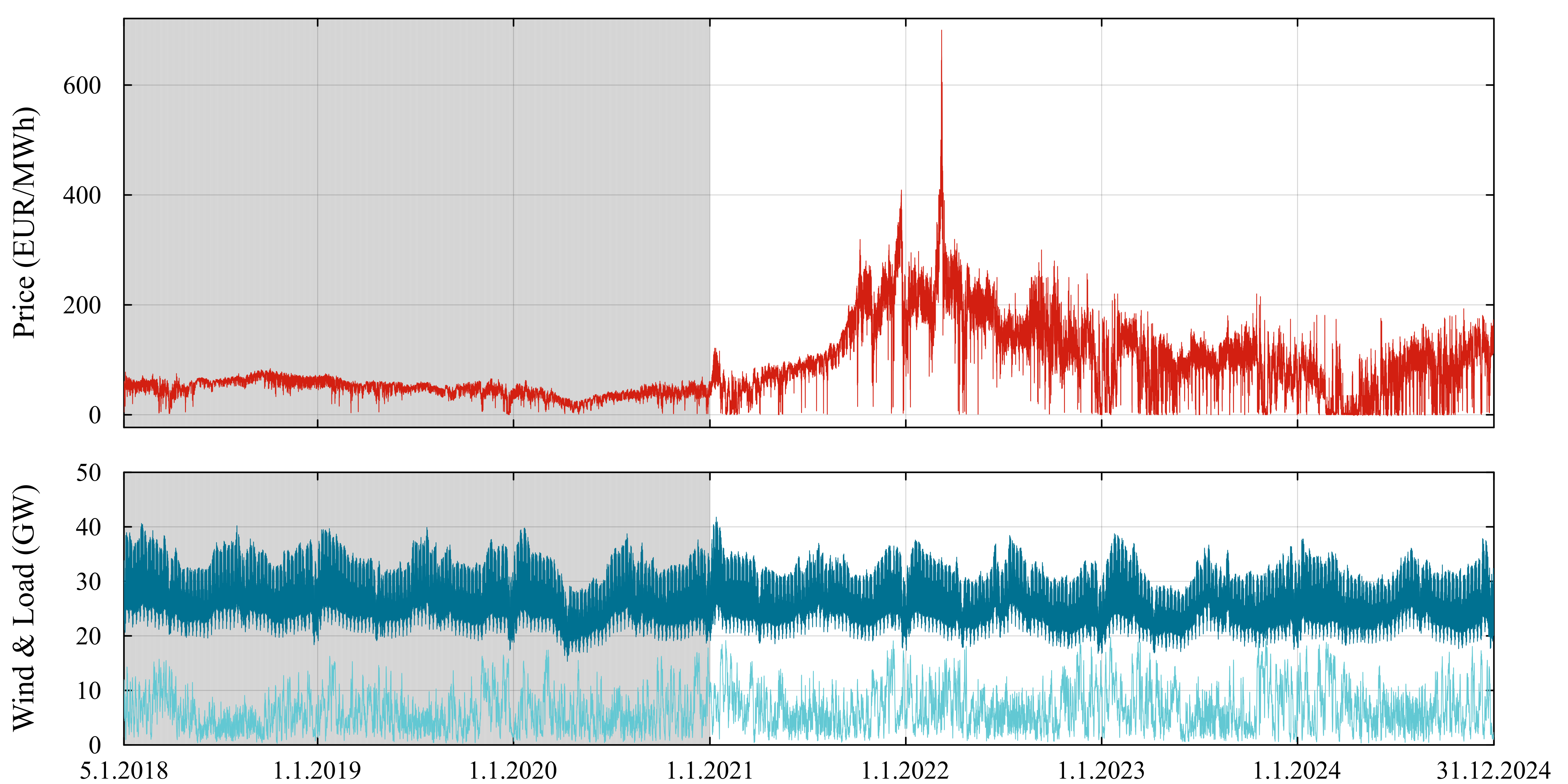}
    \end{minipage}
    \caption{German EPEX-DE (\textit{top panels}) and Spanish OMIE (\textit{bottom panels}) electricity market data: day-ahead prices (\textit{orange} or \textit{red}) and day-ahead load (\textit{dark blue}) and wind generation (\textit{light blue}) forecasts. The shaded area marks the first training window, i.e., the data used for training the models to generate forecasts for the first day in the test period (1.1.2021).}
    \label{fig:datasets}
\end{figure*}

\section{Datasets}
\label{sec:Datasets}

To ensure a sound assessment of the THieF approach, we consider two major European power markets: EPEX-DE (Germany) and OMIE (Spain). Both datasets span seven years (05.01.2018-31.12.2024, see Figure \ref{fig:datasets}) and include publicly available data:
\begin{itemize}
    \item three series specific to each market (hourly resolution) -- day-ahead prices $p_{d,h}$, day-ahead load $\hat{L}_{d, h}$ and wind $\hat{W}_{d,h}$ generation forecasts ($\hat{W}_{d,h}$ is the sum of on- and off-shore generation; source: \url{https://transparency.entsoe.eu}),
    \item two series common to both markets (daily resolution) -- the last known (hence the $d-2$ subscripts) closing prices of the nearest to delivery monthly coal $\text{API}_{d-2}$ and yearly natural gas $\text{TTF}_{d-2}$ futures (source: \url{https://www.investing.com/}). 
\end{itemize}
Note that the EPEX-DE dataset is the same as used by \cite{ser:wer:25}.

The first 1092 days (until 31.12.2020) are the initial training window. Note that after accounting for the lags in prices, see Eq.~\eqref{eqn:forecast} below, this corresponds to $1092-7=1085$ data points for training. Each day the training window is rolled forward by 24 hours. 

The remaining 4-year period (starting 01.01.2021) is a challenging test set that includes the COVID-19 pandemic (lockdowns in Germany and Spain began in March 2020, with 15.12.2020 being the first day of hard lockdown in Germany), the Russian invasion of Ukraine (24.02.2022), and the soaring natural gas prices (Q4 2021-Q4 2022). This period is also marked by the appearance of negative price spikes in Germany. For instance, on Sunday 02.07.2023 at hour 15 the price dropped to $-$500 EUR/MWh due to low demand and high renewable generation. On the other hand, negative prices were not seen in the Spanish market before 2024, and even then, prices dropped only a few EUR below zero. However, near-zero prices were reached about 5\% of the time in 2024.

\section{Computing base forecasts}
\label{sec:base:forecasts}

\subsection{Preliminaries}
\label{ssec:base:Preliminaries}

To demonstrate the versatility of our approach, we generate base forecasts using four models with four distinct architectures: linear regression, a shallow neural network, gradient boosting, and a transformer. The first two are often used as benchmarks in EPF \citep{uni:wer:zie:18,zie:wer:18,hub:mar:wer:19,mac:nit:wer:21,jan:woj:22,bil:gia:del:rav:23,uni:mac:23,ghe:zie:25,ser:wer:25}. 
The third has been reported to perform well in numerous forecasting competitions, including M5 \citep{jan:etal:22}. The fourth uses a transformer architecture, which is still rare in EPF \citep{bot:etal:23}. It is AutoGluon's new tabular foundation model called \textit{Mitra}, which excels on datasets with less than 5,000 samples and 100 features \citep{zha:rob:25}.

All four models compute the electricity price forecast $\hat{p}_{d, h}$ for day $d$ and block $h$ as a function of 20 features, as in \cite{ser:wer:25}:
\begin{align}
 \hat{p}_{d, h} = f\Big( & p_{d-1, h}, ...,\;p_{d-7, h},\;p^{min}_{d-1},\;p^{max}_{d-1}, \;\hat{L}_{d, h}, \; \nonumber \\
 & \hat{W}_{d, h}, \text{API}_{d-2}, \;\text{TTF}_{d-2}, \; D^{(1)}_d, ..., \; D^{(7)}_d \Big), 
\label{eqn:forecast}
\end{align}
where
\begin{itemize}
    \item $p_{d-i, h}$ are the lagged prices of the same block in the last seven days $i=1,...,7$,  
    \item $p_{d-1}^{min}$ and $p_{d-1}^{max}$ are the minimum and maximum hourly prices of the previous day, 
    \item $\hat{L}_{d, h}$ and $\hat{W}_{d, h}$ are the day-ahead load and wind generation forecasts for the target block, 
    \item $\text{API}_{d-2}$ and $\text{TTF}_{d-2}$ are the closing prices of the nearest to delivery monthly coal (API2) and yearly natural gas futures (TTF) from day $d-2$, and 
    \item $D_d^{(i)}$ are the weekday dummies. 
\end{itemize}
All models are trained independently for each block $h$ using a 3-year window of past values; each day the window is rolled forward by one day. Overall, we consider 60 blocks: 24 $\times$ 1H (i.e., 24 one-hour blocks), 12 $\times$ 2H, 8 $\times$ 3H, 6 $\times$ 4H, 4 $\times$ 6H, 3 $\times$ 8H, 2 $\times$ 12H, and 1 $\times$ 24H (i.e., baseload).

For more robust parameter estimation, following \cite{uni:wer:zie:18} and \cite{lag:mar:sch:wer:21}, before training the models, we preprocess the inputs using the \textit{area hyperbolic sine} transformation: $\text{asinh}\left((y - \hat{\mu}_y)/\hat{\sigma}_y\right)$, where $\hat{\mu}_y$ and $\hat{\sigma}_y$ are the sample mean and sample standard deviation of $y$ estimated on the training set. More precisely, $p_{d-i, h}$ for all $i=1,...,7$ and each selected block $h$ are transformed with a common $\hat{\mu}_y$ and $\hat{\sigma}_y$ calculated for the vector $[p_{d-1092, h}, ..., p_{d-1, h}]$. The extremes $p^{min}_{d-1}$ and $p^{max}_{d-1}$ as well as the exogenous variables $\hat{L}_{d, h}$, $\hat{W}_{d, h}$, $\text{API}_{d-2}$ and $\text{TTF}_{d-2}$ are transformed independently, also using 3-year vectors of past values; weekday dummies are not transformed.

\subsection{AutoRegression with eXogenous inputs (ARX)}
\label{ssec:ARX}
This expert -- in the sense of \cite{zie:wer:18} -- model is estimated via ordinary least squares (OLS) and uses the same inputs as \cite{ser:wer:25} to allow for direct comparisons. More precisely, the ARX model is given by:
\begin{align}
 \hat{p}_{d, h} = & \textstyle\sum_{i = 1}^{7} \beta_i p_{d-i, h} +  \beta_8 p^{min}_{d-1} + \beta_9 p^{max}_{d-1} + \beta_{10} \hat{L}_{d, h} \nonumber \\
 & + \beta_{11} \hat{W}_{d, h} + \beta_{12}\text{API}_{d-2} + \beta_{13}\text{TTF}_{d-2} \nonumber \\ 
 & + \textstyle\sum_{i=14}^{20} \beta_{i}D^{(i)}_d.
\label{eqn:arx:forecast}
\end{align}

\subsection{Nonlinear ARX (NARX)}
\label{ssec:NARX}
The nonlinear counterpart of ARX approximates $f(\cdot)$ in Eq.~\eqref{eqn:forecast} with a shallow feedforward neural network using the series-parallel architecture \citep{xie:tan:lia:09}. As in \cite{hub:mar:wer:19}, the hidden layer consists of 5 neurons and uses hyperbolic tangent activation, with a linear function in the output layer. The weights are calculated in Matlab R2025a using the Levenberg-Marquardt algorithm with early stopping based on a 10\% validation set. To mitigate the effects of parameter estimation uncertainty, we clip the network output to $[-3, 3]$ before applying the inverse transformation (i.e., the hyperbolic sine, see Section \ref{ssec:base:Preliminaries}) and use a committee machine of 10 networks \citep{mar:uni:wer:19}. More precisely, the final $\hat{p}_{d, h}$ is obtained by training the network 10 times for each $d$ and $h$, and averaging the 10 price forecasts; no hyperparameter optimization is performed.

\subsection{EXtreme Gradient Boosting (XGB)}
\label{ssec:XGB}
The third model is an ensemble of gradient-boosted decision trees \cite[GBDT;][]{has:tib:fri:09}. We use eXtreme GBDT implemented in the Python package XGBoost v3.0~\citep{che:gue:16}, with the mean squared error (MSE) as the loss function for training and hyperparameter tuning. The latter is carried out independently for each of the 60 blocks (see Section \ref{ssec:base:Preliminaries}) using Bayesian sequential optimization \citep{optuna_2019} on the initial 3-year training window (05.01.2018-31.12.2024; see Figure \ref{fig:datasets}) with a random 10\% left out for validation. We select from the following set of hyperparameters: maximum depth of a single tree (between 2 and 10), learning rate (between $10^{-4}$ and $1$ on log scale), subsample ratio of the training instances (between $\frac12$ and 1), minimum sum of instance weight required for partitioning a leaf node (between 0 and 10), minimum loss reduction required for partioning a leaf node (between 0 and $\frac12$), strengths of the L1 and L2 regularization on model weights (each between $10^{-3}$ and $10$ on log scale). Additionally, we apply early stopping to limit the number of boosting rounds (capped at 1000). Since the hyperparameter selection process is stochastic, we perform it 10 times and the final $\hat{p}_{d, h}$ is obtained by averaging the forecasts generated from 10 XGB instances with different hyperparameter sets.

\subsection{Mitra}
\label{ssec:Mitra}

The final forecaster is Amazon's state-of-the-art tabular foundation model called \textit{Mitra}, released with AutoGluon v1.4 in July 2025, based on a 12-layer 72 million-parameter transformer architecture \citep{zha:rob:25}. Since it is pre-trained on purely synthetic data, evaluating its performance on historic time series does not pose data contamination issues. Although Mitra is not specifically tailored to time series, our forecasting task requires only one-step-ahead predictions, with Eq.~\eqref{eqn:forecast} being equivalent to a tabular regression problem. 

We generate forecasts in \textit{zero-shot} mode, i.e., do not optimize the hyperparameters and rely on a pre-trained model specialized for regression tasks. We fit Mitra using AutoGluon's \textit{TabularPredictor} class with finetuning turned off and other settings kept default. Mitra is based on the in-context learning paradigm, hence the training set acts as support examples that condition the prediction. A single forecast $\hat{p}_{d, h}$ is generated by passing input data, see Eq.~\eqref{eqn:forecast}, as a query. As with the other models considered in our study, we use Mitra in a rolling window approach and train one model for each day $d$ and block $h$ in the test period.

\begin{table*}[tb]
\caption{Mean absolute errors and root mean squared errors of the base (MAE, RMSE) and reconciled ($\text{MAE}_R, \text{RMSE}_R$) forecasts with the respective gains from reconciliation (\%) for all considered hierarchy levels over the 4-year test period in the German (\textit{left}) and Spanish (\textit{right}) markets. All reconciled forecasts are significantly more accurate than the corresponding unreconciled ones, as measured by the multivariate variant of the Diebold-Mariano test \citep{zie:wer:18,lag:mar:sch:wer:21} based on the L2-norm of forecast errors; see text for details.}
\label{tab:results}
\begin{center}
\scriptsize
\begin{tabular}{cccccccccccccc}
\hline
\multicolumn{2}{l}{\textbf{}} & \multicolumn{6}{c}{\textbf{Germany (EPEX-DE)}} & \multicolumn{6}{c}{\textbf{Spain (OMIE)}} \\
\hline
\multicolumn{1}{l}{\textbf{Level}} & \multicolumn{1}{l}{\textbf{Model}} & \textbf{MAE} & $\textbf{MAE}_R$ & \textbf{\%}             & \textbf{RMSE} & $\textbf{RMSE}_R$ & \textbf{\%}              & \textbf{MAE} & $\textbf{MAE}_R$ & \textbf{\%}            & \textbf{RMSE} & $\textbf{RMSE}_R$ & \textbf{\%}            \\
\hline
                              & ARX                           & 26.93        & 26.17          & \cellcolor[HTML]{E6F3EC}2.8  & 39.83         & 38.75           & \cellcolor[HTML]{EAF5EF}2.7  & 16.72        & 16.57          & \cellcolor[HTML]{F0F7F4}0.9 & 24.40         & 24.38           & \cellcolor[HTML]{FCFCFF}0.1 \\
                              & NARX                          & 23.06        & 22.29          & \cellcolor[HTML]{DEF0E6}3.3  & 35.62         & 34.63           & \cellcolor[HTML]{E9F4EE}2.8  & 16.57        & 16.24          & \cellcolor[HTML]{D2EBDB}2.0 & 24.07         & 23.92           & \cellcolor[HTML]{EFF7F4}0.6 \\
                              & XGB                           & 23.33        & 22.30          & \cellcolor[HTML]{CEEAD7}4.4  & 37.05         & 35.02           & \cellcolor[HTML]{C7E7D1}5.5  & 16.54        & 15.98          & \cellcolor[HTML]{ADDCBB}3.4 & 24.57         & 23.67           & \cellcolor[HTML]{A4D9B3}3.7 \\
\multirow{-4}{*}{1H}          & Mitra                         & 21.37        & 21.11          & \cellcolor[HTML]{FCFCFF}1.2  & 33.31         & 32.91           & \cellcolor[HTML]{FCFCFF}1.2  & 15.92        & 15.84          & \cellcolor[HTML]{F8FBFC}0.6 & 23.61         & 23.20           & \cellcolor[HTML]{D4ECDD}1.7 \\
\hline
                              & ARX                           & 26.44        & 25.68          & \cellcolor[HTML]{E5F3EB}2.9  & 38.95         & 37.86           & \cellcolor[HTML]{E9F4EE}2.8  & 16.41        & 16.25          & \cellcolor[HTML]{EDF6F3}1.0 & 23.96         & 23.92           & \cellcolor[HTML]{FAFCFD}0.2 \\
                              & NARX                          & 22.64        & 21.61          & \cellcolor[HTML]{CDE9D6}4.5  & 34.84         & 33.44           & \cellcolor[HTML]{D9EEE1}4.0  & 16.29        & 15.80          & \cellcolor[HTML]{B8E1C4}3.0 & 23.71         & 23.29           & \cellcolor[HTML]{D3ECDC}1.8 \\
                              & XGB                           & 22.74        & 21.66          & \cellcolor[HTML]{C9E8D3}4.8  & 35.87         & 33.88           & \cellcolor[HTML]{C6E6D0}5.6  & 16.19        & 15.56          & \cellcolor[HTML]{A0D7AF}3.9 & 24.11         & 23.09           & \cellcolor[HTML]{97D3A8}4.2 \\
\multirow{-4}{*}{2H}          & Mitra                         & 20.92        & 20.38          & \cellcolor[HTML]{E9F4EE}2.6  & 32.43         & 31.60           & \cellcolor[HTML]{EBF6F1}2.6  & 15.56        & 15.38          & \cellcolor[HTML]{E8F4EE}1.1 & 23.17         & 22.54           & \cellcolor[HTML]{BCE2C8}2.7 \\
\hline
                              & ARX                           & 26.11        & 25.35          & \cellcolor[HTML]{E4F3EB}2.9  & 38.28         & 37.17           & \cellcolor[HTML]{E7F4ED}2.9  & 16.16        & 16.01          & \cellcolor[HTML]{EDF6F2}1.0 & 23.60         & 23.52           & \cellcolor[HTML]{F7FAFB}0.3 \\
                              & NARX                          & 22.12        & 21.17          & \cellcolor[HTML]{D0EAD9}4.3  & 33.89         & 32.63           & \cellcolor[HTML]{DDF0E4}3.7  & 16.01        & 15.54          & \cellcolor[HTML]{B9E1C5}2.9 & 23.44         & 22.86           & \cellcolor[HTML]{C2E5CD}2.5 \\
                              & XGB                           & 22.26        & 21.24          & \cellcolor[HTML]{CCE9D5}4.6  & 35.10         & 33.09           & \cellcolor[HTML]{C4E6CF}5.7  & 15.93        & 15.27          & \cellcolor[HTML]{97D3A8}4.2 & 23.75         & 22.67           & \cellcolor[HTML]{8FD0A1}4.5 \\
\multirow{-4}{*}{3H}          & Mitra                         & 20.61        & 19.90          & \cellcolor[HTML]{DCEFE4}3.5  & 31.87         & 30.69           & \cellcolor[HTML]{DEF0E5}3.7  & 15.31        & 15.10          & \cellcolor[HTML]{E4F2EA}1.3 & 22.72         & 22.09           & \cellcolor[HTML]{BAE2C6}2.8 \\
\hline
                              & ARX                           & 25.65        & 24.85          & \cellcolor[HTML]{E2F2E8}3.1  & 37.50         & 36.33           & \cellcolor[HTML]{E5F3EB}3.1  & 15.88        & 15.71          & \cellcolor[HTML]{EBF5F0}1.1 & 23.25         & 23.17           & \cellcolor[HTML]{F6FAFA}0.4 \\
                              & NARX                          & 21.75        & 20.66          & \cellcolor[HTML]{C6E6D0}5.0  & 33.27         & 31.75           & \cellcolor[HTML]{D2EBDB}4.6  & 15.68        & 15.27          & \cellcolor[HTML]{C2E5CD}2.6 & 22.96         & 22.48           & \cellcolor[HTML]{CCE9D6}2.1 \\
                              & XGB                           & 22.01        & 20.72          & \cellcolor[HTML]{BAE1C6}5.8  & 34.24         & 32.19           & \cellcolor[HTML]{C1E4CC}6.0  & 15.78        & 14.95          & \cellcolor[HTML]{7AC88F}5.3 & 23.60         & 22.26           & \cellcolor[HTML]{73C589}5.7 \\
\multirow{-4}{*}{4H}          & Mitra                         & 20.13        & 19.39          & \cellcolor[HTML]{D9EEE1}3.7  & 30.86         & 29.79           & \cellcolor[HTML]{E0F1E7}3.5  & 15.01        & 14.79          & \cellcolor[HTML]{E1F1E8}1.4 & 22.19         & 21.70           & \cellcolor[HTML]{C8E7D2}2.2 \\
\hline
                              & ARX                           & 25.14        & 24.35          & \cellcolor[HTML]{E1F1E8}3.1  & 36.42         & 35.30           & \cellcolor[HTML]{E5F3EB}3.1  & 15.60        & 15.45          & \cellcolor[HTML]{ECF6F1}1.0 & 22.77         & 22.64           & \cellcolor[HTML]{F1F8F6}0.6 \\
                              & NARX                          & 21.08        & 20.00          & \cellcolor[HTML]{C4E6CF}5.1  & 32.00         & 30.52           & \cellcolor[HTML]{D2EBDA}4.6  & 15.27        & 14.86          & \cellcolor[HTML]{BFE4CB}2.7 & 22.41         & 21.93           & \cellcolor[HTML]{CAE8D4}2.2 \\
                              & XGB                           & 21.52        & 20.15          & \cellcolor[HTML]{B2DEBF}6.4  & 33.38         & 31.05           & \cellcolor[HTML]{B4DFC1}7.0  & 15.41        & 14.60          & \cellcolor[HTML]{7BC890}5.3 & 22.93         & 21.74           & \cellcolor[HTML]{7FCA93}5.2 \\
\multirow{-4}{*}{6H}          & Mitra                         & 19.51        & 18.77          & \cellcolor[HTML]{D8EEE0}3.8  & 29.74         & 28.48           & \cellcolor[HTML]{D7EDDF}4.2  & 14.77        & 14.39          & \cellcolor[HTML]{C3E5CE}2.6 & 21.87         & 21.12           & \cellcolor[HTML]{ABDCB9}3.4 \\
\hline
                              & ARX                           & 24.37        & 23.65          & \cellcolor[HTML]{E4F3EA}2.9  & 35.59         & 34.35           & \cellcolor[HTML]{E0F1E7}3.5  & 15.14        & 14.99          & \cellcolor[HTML]{EDF6F2}1.0 & 22.12         & 21.99           & \cellcolor[HTML]{F0F8F5}0.6 \\
                              & NARX                          & 20.54        & 19.42          & \cellcolor[HTML]{BFE4CB}5.4  & 31.20         & 29.71           & \cellcolor[HTML]{D0EAD9}4.8  & 15.01        & 14.52          & \cellcolor[HTML]{B1DEBE}3.2 & 21.92         & 21.35           & \cellcolor[HTML]{BEE3CA}2.6 \\
                              & XGB                           & 21.06        & 19.63          & \cellcolor[HTML]{ACDCBA}6.8  & 32.94         & 30.24           & \cellcolor[HTML]{A5D9B4}8.2  & 15.10        & 14.18          & \cellcolor[HTML]{63BE7B}6.1 & 22.50         & 21.08           & \cellcolor[HTML]{63BE7B}6.3 \\
\multirow{-4}{*}{8H}          & Mitra                         & 19.45        & 18.22          & \cellcolor[HTML]{B3DFC0}6.3  & 29.34         & 27.74           & \cellcolor[HTML]{C7E7D2}5.5  & 14.44        & 14.01          & \cellcolor[HTML]{B7E0C4}3.0 & 21.45         & 20.47           & \cellcolor[HTML]{8FD0A1}4.5 \\
\hline
                              & ARX                           & 22.81        & 22.14          & \cellcolor[HTML]{E4F3EA}2.9  & 32.68         & 31.79           & \cellcolor[HTML]{EAF5EF}2.7  & 14.27        & 14.22          & \cellcolor[HTML]{FCFCFF}0.4 & 20.83         & 20.77           & \cellcolor[HTML]{F7FAFB}0.3 \\
                              & NARX                          & 19.37        & 17.80          & \cellcolor[HTML]{99D4A9}8.1  & 28.96         & 27.07           & \cellcolor[HTML]{BAE2C6}6.5  & 14.25        & 13.80          & \cellcolor[HTML]{B3DFC0}3.1 & 20.64         & 20.23           & \cellcolor[HTML]{CFEAD8}2.0 \\
                              & XGB                           & 19.70        & 17.95          & \cellcolor[HTML]{8DCF9F}8.9  & 30.55         & 27.55           & \cellcolor[HTML]{91D1A2}9.8  & 14.29        & 13.49          & \cellcolor[HTML]{71C487}5.6 & 21.24         & 19.99           & \cellcolor[HTML]{6DC284}5.9 \\
\multirow{-4}{*}{12H}         & Mitra                         & 17.59        & 16.57          & \cellcolor[HTML]{BAE2C6}5.8  & 26.27         & 24.81           & \cellcolor[HTML]{C6E6D0}5.6  & 13.67        & 13.28          & \cellcolor[HTML]{BBE2C7}2.8 & 20.16         & 19.39           & \cellcolor[HTML]{A0D7B0}3.9 \\
\hline
                              & ARX                           & 20.92        & 20.50          & \cellcolor[HTML]{F2F8F6}2.0  & 30.28         & 29.27           & \cellcolor[HTML]{E2F2E8}3.4  & 13.57        & 13.40          & \cellcolor[HTML]{E6F3EC}1.2 & 19.50         & 19.39           & \cellcolor[HTML]{F1F8F6}0.6 \\
                              & NARX                          & 17.99        & 16.53          & \cellcolor[HTML]{98D4A9}8.2  & 27.24         & 25.08           & \cellcolor[HTML]{A8DAB7}7.9  & 13.29        & 12.94          & \cellcolor[HTML]{C2E5CD}2.6 & 19.14         & 18.98           & \cellcolor[HTML]{E9F5EF}0.9 \\
                              & XGB                           & 18.80        & 16.59          & \cellcolor[HTML]{63BE7B}11.7 & 29.50         & 25.54           & \cellcolor[HTML]{63BE7B}13.4 & 13.25        & 12.58          & \cellcolor[HTML]{80CA94}5.1 & 19.61         & 18.66           & \cellcolor[HTML]{88CD9B}4.8 \\
\multirow{-4}{*}{24H}         & Mitra                         & 16.56        & 15.22          & \cellcolor[HTML]{99D4AA}8.1  & 24.81         & 22.70           & \cellcolor[HTML]{A1D7B1}8.5  & 12.91        & 12.42          & \cellcolor[HTML]{A3D8B2}3.8 & 18.84         & 18.11           & \cellcolor[HTML]{A0D7B0}3.8
\end{tabular}
\end{center}
\end{table*}

\section{Empirical results}
\label{sec:Results}

\begin{figure*}[tbp]
    \centering
    \includegraphics[width=\linewidth]{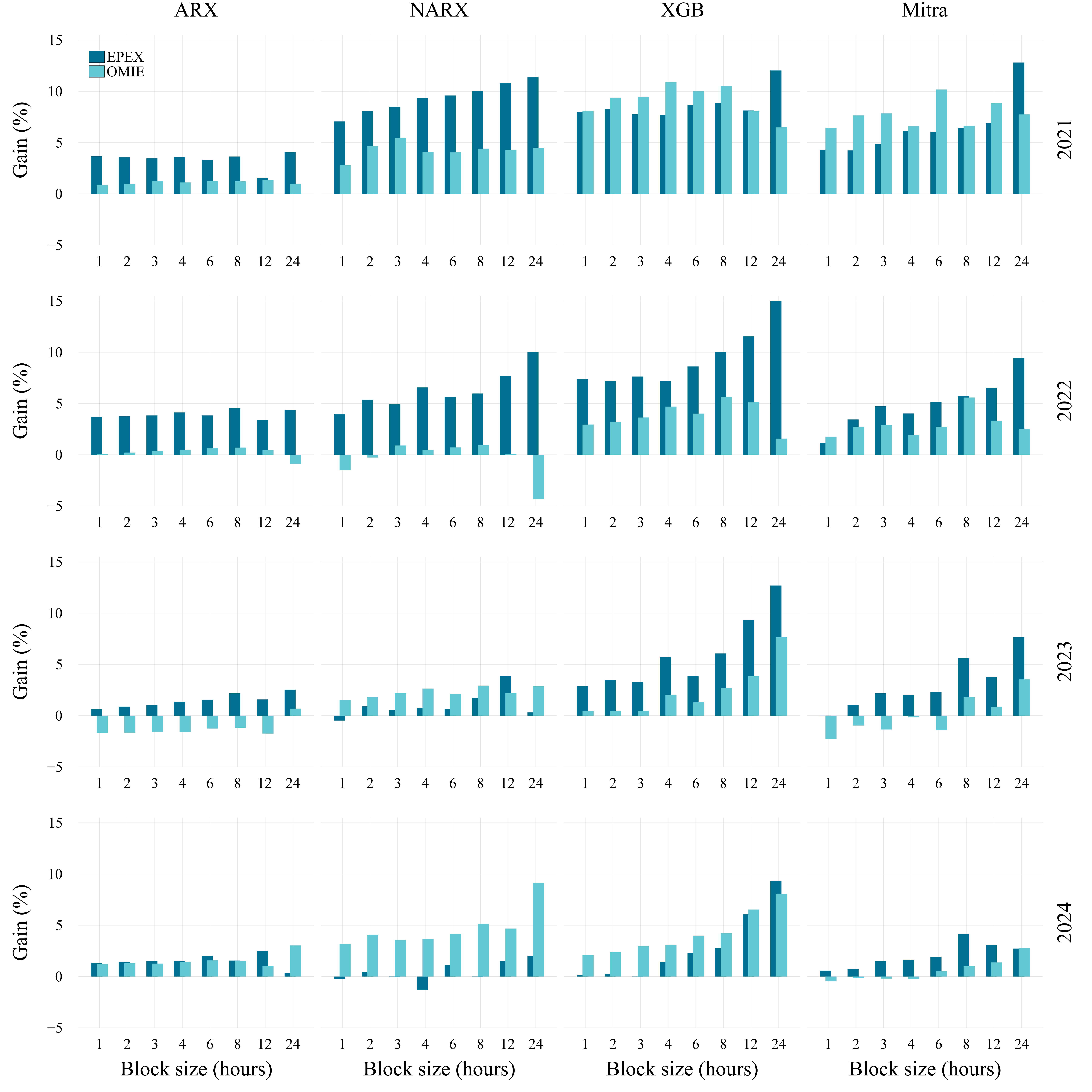}
    \caption{Percentage gains in RMSE from forecast reconciliation in the German EPEX (\textit{dark blue}) and Spanish OMIE (\textit{light blue}) markets for block sizes ranging from 1 to 24 hours, separately for each of the four models (\textit{left to right}) and calendar years (\textit{top to bottom}).}
    \label{fig:gains}
\end{figure*}

To measure the differences in predictive performance, we compute the mean absolute errors (MAE) and the root mean squared errors (RMSE) of the four models presented in Section \ref{sec:base:forecasts} for both markets and over the entire 4-year test period. The errors and the respective percentage gains from forecast reconciliation are reported in Table \ref{tab:results}. Additionally, the gains are plotted in Figure \ref{fig:gains}. Clearly, THieF improves forecast accuracy across all aggregation levels, for every considered base model and for both markets. 

The relative gains in the German (EPEX-DE) market are in general greater than in the Spanish (OMIE) market; The highest accuracy improvements in MAE are obtained for the XGB model, with a 6.7\% average gain in Germany and a 4.8\% in Spain. XGB also enjoys the highest average reconciliation gains in terms of RMSE, with improvements of 7.6\% and 5.0\%, respectively. Additionally, gains are typically higher for larger blocks, which is particularly relevant given the quadrupling of the number of products in the European day-ahead market and the resulting importance of block trading \citep{EC:25:96}.

Importantly, all reconciled forecasts are significantly more accurate than the corresponding unreconciled ones, as measured by the multivariate variant of the Diebold-Mariano test \citep{zie:wer:18,lag:mar:sch:wer:21} based on the L2-norm of forecast errors. For the EPEX-DE market, the test indicated significance at the 1\% level for each aggregation level (1H, 2H, 3H, 4H, 6H, 8H, 12H, 24H) and model. The same holds for the OMIE market with the exception of three isolated cases: 1H NARX, 1H Mitra and 24H NARX, for which the differences in unconditional predictive ability were significant at the 5\% level.

The fact that the results are consistent across architectures ranging from parsimonious linear regression with 20 parameters to a state-of-the-art transformer with up to 72 million parameters is a strong argument in favor of using the THieF approach in day-ahead EPF. 
Moreover, the computational overhead of implementing THieF in a forecasting pipeline scales linearly with the baseline. The cost of the reconciliation itself is negligible, hence the only relevant cost is estimating models at additional aggregation levels. Overall, THieF increases the computational cost to about 2.5 times that of the 24-hour baseline. This increase can be considered a minor overhead in the context of day-ahead electricity price forecasting, since the per-day runtime for a single ARX, NARX, or XGB model is sub-second on a consumer-grade CPU, whereas Mitra requires roughly one second on an NVIDIA RTX A600 GPU. 

Additionally, to analyze the reconciliation gains not only across different models but also across market regimes, Figure~\ref{fig:gains} provides a yearly breakdown of the RMSE-based gains reported in Table~\ref{tab:results}. Notably, THieF delivers accuracy gains in the German EPEX market across all years and blocks for ARX, XGB, and Mitra. For NARX, there are a few instances in which the reconciled forecasts are less accurate than the unreconciled ones, in particular for hourly forecasts in 2023-2024 and for 3-, 4-, and 8-hour blocks in 2024.

The results for the Spanish OMIE are less robust. While XGB is consistently improved by THieF in each year considered, ARX does not benefit in 2022 and loses accuracy in 2023. Similarly, the advantage of THieF is less evident for NARX in 2022 and for Mitra in 2023-2024. These findings may reflect how relevant the historical price dynamics from the 2022 energy crisis are for forecasting future prices in these two markets. Note that Spanish electricity prices exhibit lower volatility from 2022Q3 onward: after the significant spikes in the first half of 2022, prices remained below 300 EUR/MWh, with only a few spikes above 200 EUR/MWh in 2023-2024. After the extreme volatility during the crisis, the Spanish market entered a relatively calmer regime, whereas large price spikes were still observed in the EPEX market even in the final days of 2024. 

\section{Conclusions}
\label{sec:Conclusions}

In this study, we demonstrated the benefits of \textit{temporal hierarchy forecasting} (THieF) for predicting electricity prices. We showed this for the German and Spanish power markets, using a variety of base forecasting methods with increasing complexity and diverse modeling assumptions. We found a consistent and a statistically significant improvement from reconciliation in all setups -- up to 5.5\% for hourly and 13.4\% for baseload prices -- with increasing gains for higher aggregation levels ($\rightarrow$ larger blocks). 

The latter is particularly relevant given the quadrupling of the number of products \citep{EC:25:96}. On 30 September 2025, the European day-ahead electricity markets moved from hourly ($\rightarrow$ 24 products per day) to 15-minute trading intervals ($\rightarrow$ 96 products per day). Although this finer granularity offers a more accurate representation of expected generation and demand, it also adds complexity to electricity trading, thereby increasing the importance of block trading and temporal reconciliation.

In principle, if the base forecasts would have no misspecification, there would be limited space for THieF to improve accuracy. However, in practice, any forecasting method will have some degree of misspecification from the unknown underlying electricity price generating process. THieF augments these base forecasts by enabling them to better capture the different short and long-term dynamics present in the time series. An interesting finding is that the improvements are consistent for all base forecasts, including Mitra. The foundation model's misspecification is arguably of a different nature because it is used in a zero-shot, pre-trained mode that does not involve estimating model parameters. Moreover, our evaluation provides additional evidence of the usefulness of THieF when base forecasts use covariates, with the majority of applications in the literature focusing on the univariate case.

There are various extensions that stem from this work. We applied the same base forecast to all aggregation levels, however, more diverse model selection could be beneficial, better leveraging the properties of each level, and thus minimizing computational overheads. Likewise, considering a richer set of base models would be interesting, including the well-performing in EPF applications LASSO-Estimated AutoRegressions \cite[LEAR;][]{lag:mar:sch:wer:21} and Distributional Deep Neural Networks \cite[DDNN;][]{mar:nar:wer:zie:23}. Furthermore, investigating the impact of temporal reconciliation on probabilistic EPF is a useful extension. Beyond any accuracy gains, THieF provides reconciled forecasts across all levels, which can support the alignment of different decisions and potentially new strategies in energy markets. Finally, the temporal hierarchy was restricted to using aggregation levels based solely on factors of $m$. Unequal aggregation blocks are relevant for decisions supported by EPF, and embedding these in THieF can be beneficial both for forecasting performance and decisions. These are fruitful avenues for future research. 

\section*{CRediT} 

Conceptualization -- NK, AL, RW; Data curation -- AL; Formal analysis -- NK, AL; Funding acquisition -- NK, RW; Investigation -- KB, AL; Methodology -- NK, AL, RW; Software -- KB, AL; Supervision -- RW; Validation -- NK, AL, RW; Visualization -- AL; Writing (original draft) -- KB, NK, AL, RW; Writing (review \& editing) -- NK, RW.

\section*{Data and code availability}

The datasets and sources are described in Section~\ref{sec:Datasets}. The replication package is publicly available at \url{https://github.com/lipiecki/thief}.

\section*{Acknowledgments}

The study was partially supported by the National Science Center (NCN, Poland) through grants no.\ 2018/30/A/HS4/00444 (to AL and KB) and 2021/43/I/HS4/02578 (to RW), and by Riksbankens Jubileumsfond (Sweden) through grant no.\ SAB22-0073 (to NK).

\appendix

\section{Results for alternative covariance matrix approximations}\label{app:extraW}

In Table~\ref{tab:results:sch:str} we provide the results for the diagonal target shrinkage of \cite{sch:str:2005}. The results are very similar to those in Table~\ref{tab:results} for the constant correlation shrinkage of \cite{led:wol:04}. 
For the EPEX-DE market, the multivariate variant of the Diebold-Mariano test \citep{zie:wer:18,lag:mar:sch:wer:21} indicated significance at the 1\% level for each aggregation level (1H, 2H, 3H, 4H, 6H, 8H, 12H, 24H) and model. The same holds for the OMIE market with the exception of three isolated cases. However, this time the improvements in predictive ability for 1H NARX, 1H Mitra and 24H NARX are not significant at the 5\% level.

In Table~\ref{tab:results:wls} we provide the results for the so-called `variance scaling' \citep{ath:hyn:kou:pet:17} that sets all off-diagonal elements to zero and corresponds to the WLS estimator. Clearly, in this case the gains from reconciliation are smaller, and in a few cases losses in accuracy can be observed. This indicates that incorporating the correlation of errors is necessary for effective reconciliation in day-ahead EPF.

\begin{table*}[tbp]
\caption{Results for forecast reconciliation based on the diagonal target shrinkage of \cite{sch:str:2005}. Like in Table \ref{tab:results}, we report mean absolute errors and root mean squared errors of the base (MAE, RMSE) and reconciled ($\text{MAE}_R, \text{RMSE}_R$) forecasts with the respective gains from reconciliation (\%) for all considered hierarchy levels over the 4-year test period in the German (\textit{left}) and Spanish (\textit{right}) markets.}
\label{tab:results:sch:str}
\begin{center}
\scriptsize
\begin{tabular}{cccccccccccccc}
\hline
\multicolumn{2}{l}{\textbf{}} & \multicolumn{6}{c}{\textbf{Germany (EPEX-DE)}} & \multicolumn{6}{c}{\textbf{Spain (OMIE)}} \\
\hline
\multicolumn{1}{l}{\textbf{Level}} & \multicolumn{1}{l}{\textbf{Model}} & \textbf{MAE} & $\textbf{MAE}_R$ & \textbf{\%}             & \textbf{RMSE} & $\textbf{RMSE}_R$ & \textbf{\%}              & \textbf{MAE} & $\textbf{MAE}_R$ & \textbf{\%}            & \textbf{RMSE} & $\textbf{RMSE}_R$ & \textbf{\%}            \\
\hline
                              & ARX                           & 26.93                            & 26.08                              & \cellcolor[HTML]{E3F2E9}3.1  & 39.83                             & 38.56                               & \cellcolor[HTML]{E5F3EB}3.2  & 16.72                            & 16.52                              & \cellcolor[HTML]{E4F3EB}1.2 & 24.4                              & 24.28                               & \cellcolor[HTML]{F5F9F9}0.5 \\
                              & NARX                          & 23.06                            & 22.27                              & \cellcolor[HTML]{DEF0E5}3.4  & 35.62                             & 34.56                               & \cellcolor[HTML]{E8F4EE}3.0  & 16.57                            & 16.34                              & \cellcolor[HTML]{DFF1E6}1.4 & 24.07                             & 24.02                               & \cellcolor[HTML]{FCFCFF}0.2 \\
                              & XGB                           & 23.33                            & 22.36                              & \cellcolor[HTML]{D2EBDB}4.2  & 37.05                             & 35.13                               & \cellcolor[HTML]{CBE9D5}5.2  & 16.54                            & 16                                 & \cellcolor[HTML]{AFDDBC}3.2 & 24.57                             & 23.7                                & \cellcolor[HTML]{A8DAB7}3.5 \\
\multirow{-4}{*}{1H}          & MITRA                         & 21.37                            & 21.08                              & \cellcolor[HTML]{FCFCFF}1.4  & 33.31                             & 32.84                               & \cellcolor[HTML]{FCFCFF}1.4  & 15.92                            & 15.88                              & \cellcolor[HTML]{FCFCFF}0.3 & 23.61                             & 23.25                               & \cellcolor[HTML]{DBEFE3}1.5 \\
\hline
                              & ARX                           & 26.44                            & 25.6                               & \cellcolor[HTML]{E1F1E8}3.2  & 38.95                             & 37.66                               & \cellcolor[HTML]{E4F3EA}3.3  & 16.41                            & 16.2                               & \cellcolor[HTML]{E4F3EB}1.2 & 23.96                             & 23.82                               & \cellcolor[HTML]{F2F8F7}0.6 \\
                              & NARX                          & 22.64                            & 21.59                              & \cellcolor[HTML]{CCE9D6}4.6  & 34.84                             & 33.37                               & \cellcolor[HTML]{D8EEE0}4.2  & 16.29                            & 15.89                              & \cellcolor[HTML]{C4E6CF}2.4 & 23.71                             & 23.39                               & \cellcolor[HTML]{DEF0E5}1.4 \\
                              & XGB                           & 22.74                            & 21.72                              & \cellcolor[HTML]{CEE9D7}4.5  & 35.87                             & 33.99                               & \cellcolor[HTML]{CAE8D4}5.3  & 16.19                            & 15.59                              & \cellcolor[HTML]{A1D8B1}3.7 & 24.11                             & 23.13                               & \cellcolor[HTML]{99D4AA}4.1 \\
\multirow{-4}{*}{2H}          & MITRA                         & 20.92                            & 20.34                              & \cellcolor[HTML]{E7F4ED}2.8  & 32.43                             & 31.52                               & \cellcolor[HTML]{EAF5F0}2.8  & 15.56                            & 15.43                              & \cellcolor[HTML]{EFF7F4}0.8 & 23.17                             & 22.59                               & \cellcolor[HTML]{C2E5CD}2.5 \\
\hline
                              & ARX                           & 26.11                            & 25.26                              & \cellcolor[HTML]{E1F1E8}3.2  & 38.28                             & 36.98                               & \cellcolor[HTML]{E3F2E9}3.4  & 16.16                            & 15.96                              & \cellcolor[HTML]{E2F2E8}1.3 & 23.59                             & 23.43                               & \cellcolor[HTML]{F0F7F4}0.7 \\
                              & NARX                          & 22.12                            & 21.14                              & \cellcolor[HTML]{CFEAD8}4.4  & 33.89                             & 32.55                               & \cellcolor[HTML]{DBEFE2}4.0  & 16.01                            & 15.63                              & \cellcolor[HTML]{C7E7D1}2.3 & 23.44                             & 22.96                               & \cellcolor[HTML]{CCE9D6}2.1 \\
                              & XGB                           & 22.26                            & 21.3                               & \cellcolor[HTML]{D1EBDA}4.3  & 35.1                              & 33.2                                & \cellcolor[HTML]{C9E7D3}5.4  & 15.93                            & 15.29                              & \cellcolor[HTML]{99D4AA}4.0 & 23.75                             & 22.71                               & \cellcolor[HTML]{91D1A3}4.4 \\
\multirow{-4}{*}{3H}          & MITRA                         & 20.61                            & 19.86                              & \cellcolor[HTML]{DAEEE1}3.7  & 31.87                             & 30.61                               & \cellcolor[HTML]{DCEFE4}3.9  & 15.31                            & 15.14                              & \cellcolor[HTML]{E7F4ED}1.1 & 22.72                             & 22.14                               & \cellcolor[HTML]{BFE4CB}2.6 \\
\hline
                              & ARX                           & 25.65                            & 24.76                              & \cellcolor[HTML]{DEF0E5}3.4  & 37.5                              & 36.13                               & \cellcolor[HTML]{E0F1E7}3.6  & 15.88                            & 15.67                              & \cellcolor[HTML]{E2F2E8}1.3 & 23.25                             & 23.07                               & \cellcolor[HTML]{EDF6F2}0.8 \\
                              & NARX                          & 21.75                            & 20.6                               & \cellcolor[HTML]{C1E5CD}5.3  & 33.27                             & 31.67                               & \cellcolor[HTML]{D0EBD9}4.8  & 15.68                            & 15.36                              & \cellcolor[HTML]{CCE9D6}2.1 & 22.96                             & 22.57                               & \cellcolor[HTML]{D6EDDE}1.7 \\
                              & XGB                           & 22.01                            & 20.78                              & \cellcolor[HTML]{BDE3C9}5.6  & 34.24                             & 32.29                               & \cellcolor[HTML]{C5E6CF}5.7  & 15.78                            & 14.97                              & \cellcolor[HTML]{79C78E}5.2 & 23.6                              & 22.29                               & \cellcolor[HTML]{75C68B}5.5 \\
\multirow{-4}{*}{4H}          & MITRA                         & 20.13                            & 19.34                              & \cellcolor[HTML]{D7EDDF}3.9  & 30.86                             & 29.7                                & \cellcolor[HTML]{DDF0E5}3.8  & 15.01                            & 14.84                              & \cellcolor[HTML]{E7F4ED}1.1 & 22.19                             & 21.75                               & \cellcolor[HTML]{CFEAD8}2.0 \\
\hline
                              & ARX                           & 25.14                            & 24.28                              & \cellcolor[HTML]{DEF0E5}3.4  & 36.42                             & 35.12                               & \cellcolor[HTML]{E0F1E7}3.6  & 15.6                             & 15.39                              & \cellcolor[HTML]{E2F2E8}1.3 & 22.77                             & 22.55                               & \cellcolor[HTML]{E8F4EE}1.0 \\
                              & NARX                          & 21.08                            & 19.94                              & \cellcolor[HTML]{C0E4CB}5.4  & 32                                & 30.42                               & \cellcolor[HTML]{CFEAD8}4.9  & 15.27                            & 14.95                              & \cellcolor[HTML]{CCE9D6}2.1 & 22.41                             & 22.02                               & \cellcolor[HTML]{D4ECDC}1.8 \\
                              & XGB                           & 21.52                            & 20.2                               & \cellcolor[HTML]{B4DFC1}6.2  & 33.38                             & 31.15                               & \cellcolor[HTML]{B8E1C4}6.7  & 15.41                            & 14.61                              & \cellcolor[HTML]{79C78E}5.2 & 22.93                             & 21.77                               & \cellcolor[HTML]{82CB96}5.0 \\
\multirow{-4}{*}{6H}          & MITRA                         & 19.51                            & 18.71                              & \cellcolor[HTML]{D4ECDC}4.1  & 29.73                             & 28.38                               & \cellcolor[HTML]{D3ECDC}4.6  & 14.77                            & 14.44                              & \cellcolor[HTML]{C9E8D3}2.2 & 21.87                             & 21.18                               & \cellcolor[HTML]{B3DFC0}3.1 \\
\hline
                              & ARX                           & 24.37                            & 23.59                              & \cellcolor[HTML]{E1F1E8}3.2  & 35.59                             & 34.15                               & \cellcolor[HTML]{DBEFE2}4.0  & 15.14                            & 14.95                              & \cellcolor[HTML]{E2F2E8}1.3 & 22.12                             & 21.9                                & \cellcolor[HTML]{E8F4EE}1.0 \\
                              & NARX                          & 20.54                            & 19.39                              & \cellcolor[HTML]{BDE3C9}5.6  & 31.2                              & 29.61                               & \cellcolor[HTML]{CDE9D6}5.1  & 15                               & 14.61                              & \cellcolor[HTML]{BFE3CA}2.6 & 21.92                             & 21.42                               & \cellcolor[HTML]{C7E7D1}2.3 \\
                              & XGB                           & 21.06                            & 19.68                              & \cellcolor[HTML]{AEDDBC}6.6  & 32.94                             & 30.34                               & \cellcolor[HTML]{A8DAB7}7.9  & 15.1                             & 14.2                               & \cellcolor[HTML]{63BE7B}6.0 & 22.5                              & 21.12                               & \cellcolor[HTML]{63BE7B}6.2 \\
\multirow{-4}{*}{8H}          & MITRA                         & 19.45                            & 18.18                              & \cellcolor[HTML]{AFDDBD}6.5  & 29.34                             & 27.63                               & \cellcolor[HTML]{C3E5CE}5.8  & 14.44                            & 14.06                              & \cellcolor[HTML]{BFE3CA}2.6 & 21.45                             & 20.53                               & \cellcolor[HTML]{94D2A5}4.3 \\
\hline
                              & ARX                           & 22.81                            & 22.1                               & \cellcolor[HTML]{E3F2E9}3.1  & 32.68                             & 31.64                               & \cellcolor[HTML]{E5F3EB}3.2  & 14.27                            & 14.19                              & \cellcolor[HTML]{F4F9F9}0.6 & 20.83                             & 20.68                               & \cellcolor[HTML]{F0F7F4}0.7 \\
                              & NARX                          & 19.37                            & 17.72                              & \cellcolor[HTML]{91D1A3}8.5  & 28.95                             & 26.94                               & \cellcolor[HTML]{B4DFC1}7.0  & 14.25                            & 13.87                              & \cellcolor[HTML]{BCE2C8}2.7 & 20.64                             & 20.28                               & \cellcolor[HTML]{D6EDDE}1.7 \\
                              & XGB                           & 19.7                             & 17.98                              & \cellcolor[HTML]{8ED0A0}8.7  & 30.55                             & 27.61                               & \cellcolor[HTML]{92D1A4}9.6  & 14.29                            & 13.5                               & \cellcolor[HTML]{71C487}5.5 & 21.24                             & 20.01                               & \cellcolor[HTML]{6EC384}5.8 \\
\multirow{-4}{*}{12H}         & MITRA                         & 17.59                            & 16.5                               & \cellcolor[HTML]{B4DFC1}6.2  & 26.27                             & 24.69                               & \cellcolor[HTML]{C1E4CC}6.0  & 13.67                            & 13.34                              & \cellcolor[HTML]{C4E6CF}2.4 & 20.16                             & 19.44                               & \cellcolor[HTML]{A6D9B5}3.6 \\
\hline
                              & ARX                           & 20.92                            & 20.52                              & \cellcolor[HTML]{F5F9F9}1.9  & 30.28                             & 29.15                               & \cellcolor[HTML]{DDF0E5}3.8  & 13.57                            & 13.36                              & \cellcolor[HTML]{DCEFE4}1.5 & 19.5                              & 19.31                               & \cellcolor[HTML]{E8F4EE}1.0 \\
                              & NARX                          & 17.99                            & 16.41                              & \cellcolor[HTML]{8CCF9F}8.8  & 27.24                             & 24.9                                & \cellcolor[HTML]{9FD7AF}8.6  & 13.29                            & 13                                 & \cellcolor[HTML]{CCE9D6}2.1 & 19.14                             & 19.01                               & \cellcolor[HTML]{F0F7F4}0.7 \\
                              & XGB                           & 18.8                             & 16.63                              & \cellcolor[HTML]{63BE7B}11.5 & 29.5                              & 25.6                                & \cellcolor[HTML]{63BE7B}13.2 & 13.25                            & 12.57                              & \cellcolor[HTML]{7CC890}5.1 & 19.61                             & 18.66                               & \cellcolor[HTML]{87CD9A}4.8 \\
\multirow{-4}{*}{24H}         & MITRA                         & 16.56                            & 15.15                              & \cellcolor[HTML]{91D1A3}8.5  & 24.81                             & 22.56                               & \cellcolor[HTML]{99D4A9}9.1  & 12.91                            & 12.43                              & \cellcolor[HTML]{A1D8B1}3.7 & 18.84                             & 18.17                               & \cellcolor[HTML]{A6D9B5}3.6
\end{tabular}
\end{center}
\end{table*}

\begin{table*}[tbp]
\caption{Results for the so-called `variance scaling' \citep{ath:hyn:kou:pet:17} that sets all off-diagonal elements to zero and corresponds to the WLS estimator. Like in Tables \ref{tab:results} and \ref{tab:results:sch:str}, we report mean absolute errors and root mean squared errors of the base (MAE, RMSE) and reconciled ($\text{MAE}_R, \text{RMSE}_R$) forecasts with the respective gains from reconciliation (\%) for all considered hierarchy levels over the 4-year test period in the German (\textit{left}) and Spanish (\textit{right}) markets.}
\label{tab:results:wls}
\begin{center}
\scriptsize
\begin{tabular}{cccccccccccccc}
\hline
\multicolumn{2}{l}{\textbf{}} & \multicolumn{6}{c}{\textbf{Germany (EPEX-DE)}} & \multicolumn{6}{c}{\textbf{Spain (OMIE)}} \\
\hline
\multicolumn{1}{l}{\textbf{Level}} & \multicolumn{1}{l}{\textbf{Model}} & \textbf{MAE} & $\textbf{MAE}_R$ & \textbf{\%}             & \textbf{RMSE} & $\textbf{RMSE}_R$ & \textbf{\%}              & \textbf{MAE} & $\textbf{MAE}_R$ & \textbf{\%}            & \textbf{RMSE} & $\textbf{RMSE}_R$ & \textbf{\%}            \\
\hline
                              & ARX                           & 26.93                            & 26.92                              & \cellcolor[HTML]{FCFCFF}0.0  & 39.83                             & 39.82                               & \cellcolor[HTML]{FCFCFF}0.0  & 16.72                            & 16.72                              & \cellcolor[HTML]{FCFCFF}0.0 & 24.4                              & 24.43                               & \cellcolor[HTML]{FBE9EC}-0.1 \\
                              & NARX                          & 23.06                            & 22.99                              & \cellcolor[HTML]{F6FAFA}0.3  & 35.62                             & 35.55                               & \cellcolor[HTML]{F8FBFC}0.2  & 16.57                            & 16.49                              & \cellcolor[HTML]{EAF5F0}0.5 & 24.07                             & 24.03                               & \cellcolor[HTML]{F3F9F7}0.2  \\
                              & XGB                           & 23.33                            & 23.25                              & \cellcolor[HTML]{F6FAFA}0.3  & 37.05                             & 36.82                               & \cellcolor[HTML]{F0F8F5}0.6  & 16.54                            & 16.51                              & \cellcolor[HTML]{F5FAF9}0.2 & 24.57                             & 24.53                               & \cellcolor[HTML]{F3F9F7}0.2  \\
\multirow{-4}{*}{1H}          & MITRA                         & 21.37                            & 21.42                              & \cellcolor[HTML]{FAB2B5}-0.2 & 33.31                             & 33.38                               & \cellcolor[HTML]{F8696B}-0.2 & 15.92                            & 15.88                              & \cellcolor[HTML]{F2F8F6}0.3 & 23.61                             & 23.54                               & \cellcolor[HTML]{EFF7F3}0.3  \\
\hline
                              & ARX                           & 26.44                            & 26.42                              & \cellcolor[HTML]{FAFCFE}0.1  & 38.95                             & 38.93                               & \cellcolor[HTML]{FCFCFF}0.0  & 16.41                            & 16.4                               & \cellcolor[HTML]{FCFCFF}0.0 & 23.96                             & 23.96                               & \cellcolor[HTML]{FCFCFF}0.0  \\
                              & NARX                          & 22.64                            & 22.29                              & \cellcolor[HTML]{DDF0E4}1.5  & 34.84                             & 34.36                               & \cellcolor[HTML]{E0F1E7}1.4  & 16.29                            & 16.06                              & \cellcolor[HTML]{CAE8D4}1.4 & 23.71                             & 23.39                               & \cellcolor[HTML]{BCE2C7}1.4  \\
                              & XGB                           & 22.74                            & 22.63                              & \cellcolor[HTML]{F2F8F6}0.5  & 35.87                             & 35.73                               & \cellcolor[HTML]{F4F9F9}0.4  & 16.19                            & 16.1                               & \cellcolor[HTML]{E7F4ED}0.6 & 24.11                             & 23.96                               & \cellcolor[HTML]{E1F1E7}0.6  \\
\multirow{-4}{*}{2H}          & MITRA                         & 20.92                            & 20.64                              & \cellcolor[HTML]{E1F1E8}1.3  & 32.43                             & 32.01                               & \cellcolor[HTML]{E2F2E9}1.3  & 15.56                            & 15.39                              & \cellcolor[HTML]{D4ECDD}1.1 & 23.17                             & 22.84                               & \cellcolor[HTML]{BCE2C7}1.4  \\
\hline
                              & ARX                           & 26.11                            & 26.07                              & \cellcolor[HTML]{FAFCFE}0.1  & 38.28                             & 38.24                               & \cellcolor[HTML]{FAFCFE}0.1  & 16.16                            & 16.15                              & \cellcolor[HTML]{F9FBFC}0.1 & 23.59                             & 23.57                               & \cellcolor[HTML]{F8FBFB}0.1  \\
                              & NARX                          & 22.12                            & 21.88                              & \cellcolor[HTML]{E5F3EC}1.1  & 33.89                             & 33.55                               & \cellcolor[HTML]{E8F4EE}1.0  & 16.01                            & 15.79                              & \cellcolor[HTML]{CAE8D4}1.4 & 23.44                             & 22.96                               & \cellcolor[HTML]{A0D7AF}2.0  \\
                              & XGB                           & 22.26                            & 22.21                              & \cellcolor[HTML]{F8FBFC}0.2  & 35.1                              & 34.97                               & \cellcolor[HTML]{F4F9F9}0.4  & 15.93                            & 15.8                               & \cellcolor[HTML]{DFF1E6}0.8 & 23.75                             & 23.56                               & \cellcolor[HTML]{D7EDDF}0.8  \\
\multirow{-4}{*}{3H}          & MITRA                         & 20.61                            & 20.19                              & \cellcolor[HTML]{D0EBDA}2.1  & 31.87                             & 31.09                               & \cellcolor[HTML]{CCE9D6}2.4  & 15.31                            & 15.09                              & \cellcolor[HTML]{CAE8D4}1.4 & 22.72                             & 22.38                               & \cellcolor[HTML]{B7E0C3}1.5  \\
\hline
                              & ARX                           & 25.65                            & 25.59                              & \cellcolor[HTML]{F8FBFC}0.2  & 37.5                              & 37.41                               & \cellcolor[HTML]{F8FBFC}0.2  & 15.88                            & 15.86                              & \cellcolor[HTML]{F9FBFC}0.1 & 23.25                             & 23.21                               & \cellcolor[HTML]{F3F9F7}0.2  \\
                              & NARX                          & 21.75                            & 21.37                              & \cellcolor[HTML]{D9EEE1}1.7  & 33.27                             & 32.68                               & \cellcolor[HTML]{D8EEE0}1.8  & 15.68                            & 15.49                              & \cellcolor[HTML]{D1EBDA}1.2 & 22.96                             & 22.58                               & \cellcolor[HTML]{B2DEBF}1.6  \\
                              & XGB                           & 22.01                            & 21.74                              & \cellcolor[HTML]{E3F2EA}1.2  & 34.24                             & 34.13                               & \cellcolor[HTML]{F6FAFA}0.3  & 15.78                            & 15.5                               & \cellcolor[HTML]{BBE2C7}1.8 & 23.6                              & 23.17                               & \cellcolor[HTML]{A9DBB7}1.8  \\
\multirow{-4}{*}{4H}          & MITRA                         & 20.13                            & 19.64                              & \cellcolor[HTML]{CAE8D4}2.4  & 30.86                             & 30.16                               & \cellcolor[HTML]{CEEAD8}2.3  & 15.01                            & 14.8                               & \cellcolor[HTML]{CAE8D4}1.4 & 22.19                             & 21.99                               & \cellcolor[HTML]{D3ECDB}0.9  \\
\hline
                              & ARX                           & 25.14                            & 25.05                              & \cellcolor[HTML]{F4F9F8}0.4  & 36.42                             & 36.3                                & \cellcolor[HTML]{F6FAFA}0.3  & 15.6                             & 15.54                              & \cellcolor[HTML]{EEF7F3}0.4 & 22.77                             & 22.67                               & \cellcolor[HTML]{E5F3EB}0.5  \\
                              & NARX                          & 21.08                            & 20.79                              & \cellcolor[HTML]{DFF1E6}1.4  & 32                                & 31.45                               & \cellcolor[HTML]{DAEFE2}1.7  & 15.27                            & 15.13                              & \cellcolor[HTML]{D8EEE0}1.0 & 22.41                             & 22.03                               & \cellcolor[HTML]{AEDDBB}1.7  \\
                              & XGB                           & 21.52                            & 21.18                              & \cellcolor[HTML]{DBEFE3}1.6  & 33.38                             & 33                                  & \cellcolor[HTML]{E6F4EC}1.1  & 15.41                            & 15.16                              & \cellcolor[HTML]{C2E5CD}1.6 & 22.93                             & 22.66                               & \cellcolor[HTML]{C5E6CF}1.2  \\
\multirow{-4}{*}{6H}          & MITRA                         & 19.51                            & 18.99                              & \cellcolor[HTML]{C4E6CF}2.7  & 29.73                             & 28.84                               & \cellcolor[HTML]{C0E4CB}3.0  & 14.77                            & 14.39                              & \cellcolor[HTML]{A1D8B1}2.5 & 21.87                             & 21.39                               & \cellcolor[HTML]{96D3A7}2.2  \\
\hline
                              & ARX                           & 24.37                            & 24.33                              & \cellcolor[HTML]{F8FBFC}0.2  & 35.59                             & 35.44                               & \cellcolor[HTML]{F4F9F9}0.4  & 15.14                            & 15.07                              & \cellcolor[HTML]{EEF7F3}0.4 & 22.12                             & 22.01                               & \cellcolor[HTML]{E5F3EB}0.5  \\
                              & NARX                          & 20.54                            & 20.16                              & \cellcolor[HTML]{D7EDDF}1.8  & 31.2                              & 30.66                               & \cellcolor[HTML]{D8EEE0}1.8  & 15                               & 14.77                              & \cellcolor[HTML]{C6E6D0}1.5 & 21.92                             & 21.45                               & \cellcolor[HTML]{9BD5AB}2.1  \\
                              & XGB                           & 21.06                            & 20.62                              & \cellcolor[HTML]{D0EBDA}2.1  & 32.94                             & 32.25                               & \cellcolor[HTML]{D2EBDB}2.1  & 15.1                             & 14.74                              & \cellcolor[HTML]{A5D9B4}2.4 & 22.5                              & 22.04                               & \cellcolor[HTML]{A0D7AF}2.0  \\
\multirow{-4}{*}{8H}          & MITRA                         & 19.45                            & 18.47                              & \cellcolor[HTML]{94D2A5}5.0  & 29.34                             & 28.07                               & \cellcolor[HTML]{A6D9B5}4.3  & 14.44                            & 13.97                              & \cellcolor[HTML]{88CD9B}3.2 & 21.45                             & 20.74                               & \cellcolor[HTML]{63BE7B}3.3  \\
\hline
                              & ARX                           & 22.81                            & 22.79                              & \cellcolor[HTML]{FAFCFE}0.1  & 32.68                             & 32.72                               & \cellcolor[HTML]{FAB2B5}-0.1 & 14.27                            & 14.26                              & \cellcolor[HTML]{F9FBFC}0.1 & 20.83                             & 20.75                               & \cellcolor[HTML]{EAF5EF}0.4  \\
                              & NARX                          & 19.37                            & 18.66                              & \cellcolor[HTML]{AFDDBD}3.7  & 28.95                             & 28.03                               & \cellcolor[HTML]{BCE2C8}3.2  & 14.25                            & 14.06                              & \cellcolor[HTML]{CDE9D7}1.3 & 20.64                             & 20.35                               & \cellcolor[HTML]{BCE2C7}1.4  \\
                              & XGB                           & 19.7                             & 19.08                              & \cellcolor[HTML]{B9E1C6}3.2  & 30.55                             & 29.8                                & \cellcolor[HTML]{CCE9D6}2.4  & 14.29                            & 14.05                              & \cellcolor[HTML]{BFE3CA}1.7 & 21.24                             & 20.99                               & \cellcolor[HTML]{C5E6CF}1.2  \\
\multirow{-4}{*}{12H}         & MITRA                         & 17.59                            & 16.74                              & \cellcolor[HTML]{98D4A9}4.8  & 26.27                             & 25.04                               & \cellcolor[HTML]{9ED6AE}4.7  & 13.67                            & 13.24                              & \cellcolor[HTML]{88CD9B}3.2 & 20.16                             & 19.63                               & \cellcolor[HTML]{84CC97}2.6  \\
\hline
                              & ARX                           & 20.92                            & 21                                 & \cellcolor[HTML]{F8696B}-0.4 & 30.28                             & 30.12                               & \cellcolor[HTML]{F0F8F5}0.6  & 13.57                            & 13.42                              & \cellcolor[HTML]{D4ECDD}1.1 & 19.5                              & 19.34                               & \cellcolor[HTML]{D7EDDF}0.8  \\
                              & NARX                          & 17.99                            & 17.37                              & \cellcolor[HTML]{B3DFC0}3.5  & 27.24                             & 26.13                               & \cellcolor[HTML]{AADBB8}4.1  & 13.29                            & 13.19                              & \cellcolor[HTML]{E3F2E9}0.7 & 19.14                             & 19.1                                & \cellcolor[HTML]{F3F9F7}0.2  \\
                              & XGB                           & 18.8                             & 17.77                              & \cellcolor[HTML]{8BCF9E}5.4  & 29.5                              & 27.93                               & \cellcolor[HTML]{92D1A3}5.3  & 13.25                            & 13.21                              & \cellcolor[HTML]{F2F8F6}0.3 & 19.61                             & 19.76                               & \cellcolor[HTML]{F8696B}-0.8 \\
\multirow{-4}{*}{24H}         & MITRA                         & 16.56                            & 15.36                              & \cellcolor[HTML]{63BE7B}7.3  & 24.81                             & 22.92                               & \cellcolor[HTML]{63BE7B}7.6  & 12.91                            & 12.36                              & \cellcolor[HTML]{63BE7B}4.2 & 18.84                             & 18.34                               & \cellcolor[HTML]{7FCA93}2.7 

\end{tabular}
\end{center}
\end{table*}

\bibliographystyle{elsarticle-harv} 
\begin{footnotesize}
\bibliography{epf}
\end{footnotesize}

\end{document}